\documentclass[ aip,
 amsmath,amssymb,
reprint,
]{revtex4-2}

\usepackage{graphicx}
\usepackage{dcolumn}
\usepackage{bm}
\usepackage[utf8]{inputenc}
\usepackage[T1]{fontenc}
\usepackage{mathptmx}
\usepackage{xcolor}
\usepackage{graphicx}
\usepackage{caption}
\usepackage{subcaption}
\usepackage{tabularray}
\usepackage{comment}
\usepackage[hmargin=1cm]{geometry}
\usepackage{ulem}
\usepackage{booktabs}
\newcommand{\quasi}[1]{{\it quasi}}

\begin{document}

\title{Spin-State Gaps and Self-Interaction-Corrected Density Functional Approximations:
Octahedral Fe(II) Complexes as Case Study}

\author{Selim Romero}
\affiliation{Computational Science Program,The University of Texas at El Paso, El Paso, Texas 79968}
\author{Tunna Baruah}
\author{Rajendra R. Zope}
\affiliation{Department of Physics, University of Texas at El Paso,TX, 79968}
\date{\today}

\begin{abstract}
  Accurate prediction of spin-state energy difference is crucial for understanding 
  the spin crossover (SCO) phenomena and is very challenging for the density 
  functional approximations, especially for the local and semi-local approximations,
  due to delocalization errors.
  Here, we investigate the effect of self-interaction error removal
  from the local spin density approximation (LSDA) and PBE generalized gradient approximation 
  on the  spin-state gaps  of Fe(II) complexes with various ligands 
  using recently developed locally scaled self-interaction correction 
  (LSIC) by Zope {\it et al.}  [J. Chem. Phys. {\bf 151}, 214108 (2019)]. The LSIC method 
  is exact for one-electron density, recovers uniform electron gas limit of 
  the underlying functional,  and approaches the well known 
  Perdew-Zunger self-interaction correction 
  (PZSIC) as a special case when scaling factor is set to unity.
  Our results, when compared with reference diffusion Monte Carlo (DMC) results, 
  show that PZSIC method significantly overestimates spin-state gaps 
  favoring low spin states for all ligands and does not improve upon DFAs.
   The perturbative LSIC-LSDA using PZSIC densities
  significantly improves the gaps  with mean absolute error of 0.51 eV but slightly
  ovecorrects for the stronger CO ligands. The quasi self-consistent LSIC-LSDA, 
  like CCSD(T), gives correct sign of spin-state gaps for all ligands with MAE of 0.56 eV,
  comparable to that of CCSD(T) (0.49 eV).

\end{abstract}
\maketitle

\section{\label{sec:level1}INTRODUCTION}
The Kohn-Sham (KS) formulation of the density functional theory (DFT) is
an exact theory that is widely used in many areas, such as
chemical physics, materials science, and condensed matter physics\cite{jones2015density}. 
Its practical usage requires approximations to the exchange-correlation energy functional, 
whose complexity determines the accuracy and efficiency of the calculations and system 
sizes that could be studied. 
Since there is no systematic way to improve upon the accuracy of exchange-correlation 
approximations, a large number of density functional approximations (DFAs) with various 
ingredients have been  proposed\cite{perdew2001jacob,MARQUES20122272}. 
Semi-local density functionals, in general,  offer a good balance of accuracy 
and efficiency, hence they are widely used in calculations for large system sizes.
While many DFAs perform well for closed-shell systems, they can 
fail to accurately describe the spin states of the open-shell systems,
such as  transition metal complexes or  organic radicals. Such systems typically have multiple electronic configurations that are close in energy thereby resulting in several accessible spin states.
Prototype  examples of such systems are spin-crossover 
Fe-centered complexes\cite{WOS:000806610000021, wilbraham2018communication, radon2019benchmarking, kulik2020making, nandy2020large, alipour2020appraising, vela2020thermal, mariano2020biased, mariano2021improved, floser2020detailed, song2018benchmarks, patra2019rethinking, droghetti2012assessment, ganzenmuller2005comparison, bowman2012low, fouqueau2004comparison, fouqueau2005comparison, pierloot2006relative, pierloot2008relative, wilbraham2017multiconfiguration, wilbraham2018communication,  verma2017assessment, ye2010accurate, ioannidis2015towards, rudavskyi2014computational, radon2019benchmarking, fumanal2016DMC, pinter2018conceptual, mortensen2015spin, lawson2012accurate, kepenekian2009energetics, domingo2010spin, janesko2017reducing, swart2008accurate}. 
In these systems,
the spin state can change with small variations in temperature.
Difficulties of the DFAs in properly describing 
the $d$ electrons in these complexes arise due to inherent 
self-interaction-error (SIE) in these functionals. SIE
can limit their ability to 
provide qualitatively accurate description of spin-state
ordering in these systems.
 
The SIE in DFA arises due to the incomplete cancellation of the self-Coulomb energy by the self-exchange energy of the 
approximate density functionals for one-electron density. 
 The consequences are poor performance for many properties,
such as, low reaction barrier heights, underestimation of eigenvalues of valence orbitals, unbound atomic anions,
overestimation of polarizabilities of molecular chains, underestimation of band gaps,
etc. \cite{perdew1985density,PhysRevLett.51.1888,patchkovskii2003improving,grafenstein2004impact,ruzsinszky2006spurious}
 Perdew and coworkers have shown that the density functional 
total energy of an $N$-electron system should vary linearly between integer numbers of electrons. 
However, with the approximate density functionals, the total energy of an $N$-electron system varies 
as a convex curve as the electronic charge varies between $N$ and $N+1$ 
electrons\cite{PhysRevLett.100.146401,cohen2008insights}.

The DFAs artificially lower the energy of a fractional electron system and thereby produce a 
convex curve instead of a linear curve between integer electron numbers. 
This deviation from the linearity is often called {\it delocalization error}
\cite{Weitao2012,WOS:000296521200011,li2018localized,WOS:000327712800019}
or
sometimes as the {\it many-electron} SIE\cite{ruzsinszky2006spurious}. 
The delocalization error is particularly severe for systems with $d$ or $f$ electrons. The SIE can result 
in incorrect charge states and consequently yield erroneous spin state ordering. 
Indeed, elimination of delocalization errors in the practical density 
functional calculations is considered as the most 
outstanding challenge 
in density functional theory\cite{ErinJohnson2022}.

 In this work, we investigate and compare the effect of one-electron self-interaction correction\cite{lindgren1971statistical}
 (SIC) on the  energy gap between spin-states of single Fe-center complexes using 
 recent locally scaled self-interaction correction (LSIC) method of Zope and coworkers\cite{zope2019LSIC} 
 and compare the results with  the well-known Perdew-Zunger SIC (PZSIC)\cite{perdew1981self} 
 method. The SIC corrections are applied to the simplest local spin density functional (LSDA)
 and one of the most widely used generalized-gradient-approximation (GGA) of 
 Perdew-Burke-Ernzerhof (PBE)\cite{}.
 Our work shows that the PZSIC performs poorly for the spin-state gaps 
 favoring low spin states for all cases
 while  LSIC methods perform significantly better (especially quasi self-consistent LSIC-LSDA)
 with an mean absolute errors comparable to those of CCSD(T).
 
 In the following Sec.~\ref{sec:sicmethods} and \ref{sec:compdetails}, we provide brief descriptions of the methods used in this work and the computational details. The results and discussion are presented in Sec.~\ref{sec:resultsanddiscussion}.

\section{The Locally scaled self-interaction correction (LSIC method)}
\label{sec:sicmethods}

 LSIC\cite{zope2019LSIC}
 is a one-electron SIC method, that is, 
 the method is exact for one electron cases, and
 it 
  employs an orbital-by-orbital correction scheme\cite{lindgren1971statistical}
 in which  the total energy is given as
\begin{equation}\label{eq:lsic}
     E^{LSIC}[\rho_\uparrow,\rho_\downarrow] = E^{DFA}[\rho_\uparrow,\rho_\downarrow] + E^{SIC}.
\end{equation}

Here,
 \begin{equation}
    E^{SIC} = - \sum_{i\sigma}^{occ}\{ U^{LSIC}[\rho_{i\sigma}] + E^{LSIC}_{XC}[\rho_{i\sigma},0] \},
      \label{eq:lsic_esic}
 \end{equation}
 with
 \begin{equation}
    U^{LSIC} [\rho_{i\sigma}] = \frac{1}{2}\int d^3r\, \, z_\sigma(\vec{r})
    \rho_{i\sigma}(\vec{r}) \, \int d\vec{r}\,' 
    \frac{\rho_{i\sigma}(\vec{r}\,')}{|\vec{r}-\vec{r}\,'|},\\
      \label{eq:lsicCoul}
\end{equation}
\begin{equation}
     E_{XC}^{LSIC} [\rho_{i\sigma},0] = \int d^3{r}\, \, z_\sigma(\vec{r})
     \rho_{i\sigma}(\vec{r})  \,  \epsilon_{XC}^{DFA} ([\rho_{i\sigma},0],\vec{r}),
      \label{eq:lsicExc}
\end{equation}
where $\epsilon_{XC}^{DFA}$ is the exchange-correlation energy density per particle. 

  Here, $z_{\sigma}(\vec{r})$ is the local scaling factor $z_{\sigma}\vec{r})=\tau_\sigma^W(\vec{r})/\tau_\sigma(\vec{r})$, where $\tau_\sigma^W = |\nabla \rho_{\sigma}|^2/8 \rho_{\sigma}$ 
is the von Weizs\"acker kinetic energy density and 
$\tau_\sigma(\vec{r}) = \frac 1 2 \Sigma_i |\nabla \psi_{i,\sigma}|^2$ is the non-interacting kinetic energy density.

 The scaling factor $z_{\sigma}$ is an iso orbital indicator which lies between 0 and 1, indicating the nature of the  charge 
 for $z_\sigma(\vec{r})=1$ as single electron density and  for $z_\sigma(\vec{r})=0$  as the uniform electron density.
Scaling the self-interaction correction terms with $z_\sigma$ thus retains the full correction
for a one-electron density, making the theory exact in that limit, and eliminates the correction in the limit of uniform density where $E_{XC}^{DFA}$ is already exact by design. 
The LSIC method can be adapted using any suitable iso-orbital indicator similar to the 
the local hybrid functionals.

   The well known PZSIC method is a special case of LSIC method when the iso-orbital $z_\sigma$ 
   in Eq. (\ref{eq:lsicCoul}) and Eq. (\ref{eq:lsicExc})
   is set to 1. The total energy in LSIC as well in PZSIC depends 
   not only on the density but also on the specific choice
of orbitals used to represent that density. Local orbitals are derived
from the canonical orbitals through a unitary transformation and the
total energy needs to be minimized with respect to the unitary transformation. 
Both PZSIC and LSIC  total energies 
are evaluated using the Fermi-L\"owdin localized orbitals as described in the next section (Sec.~\ref{sec:flosic}).

\subsection{Fermi-L\"owdin Self-Interaction-Correction}\label{sec:flosic}
We use LSIC and PZSIC  within the Fermi-L\"owdin orbital self-interaction 
correction (FLOSIC) scheme.\cite{doi:10.1063/1.4869581} 
In the FLOSIC scheme,\cite{doi:10.1063/1.4869581} the SI correction to the total energy 
in Eq.~(\ref{eq:lsic_esic}) 
is computed using  local orbitals based on Fermi orbitals (FOs)\cite{Luken1982}.
The Fermi orbitals are obtained from the spin density matrix and spin density as
 \begin{equation}
    F_{j\sigma}(\vec{r}) = \frac{\sum_i \psi_{i\sigma}(\vec{a}_{j\sigma}) \psi_{i\sigma}(\vec{r})}{\sqrt{\rho_{\sigma} (\vec{a}_{j\sigma})}},
\end{equation}
  where $i$ and $j$ are the indices of  $i^{th}$ KS orbital and $j^{th}$ FO, respectively,
 $\sigma$ is the spin index, and $\vec{a_j}$ 
 are  position coordinates  referred to as Fermi orbital descriptors (FODs).
The FOs are normalized but not orthogonal to each other. Therefore, symmetric orthogonalization  is applied to the FOs through the L\"owdin\cite{lowdin1950non} scheme to obtain Fermi-L\"owdin orbitals (FLOs).
 
  The optimal FLOs are obtained by minimizing the $E^{PZSIC}$ with respect to the FOD positions\cite{doi:10.1063/1.4907592,PEDERSON2015153}  using a gradient optimization scheme\cite{liu1989}.
 Further details regarding the FLOSIC methodology and examples of FLOSIC calculations for various properties
 are available in Refs.~\citenum{doi:10.1063/1.4869581,
doi:10.1063/1.4907592,PEDERSON2015153,yang2017flojac,doi:10.1063/1.4947042,doi:10.1002/jcc.25586,doi:10.1063/1.5120532,diaz2021kli, diaz2021implementation,aquino2020fractional,mishra2022barrier,mishra2022coupling}.
 
 \begin{figure}
    \centering
    \begin{subfigure}[b]{0.18\textwidth}
       \centering
       \includegraphics[width=0.7\textwidth]{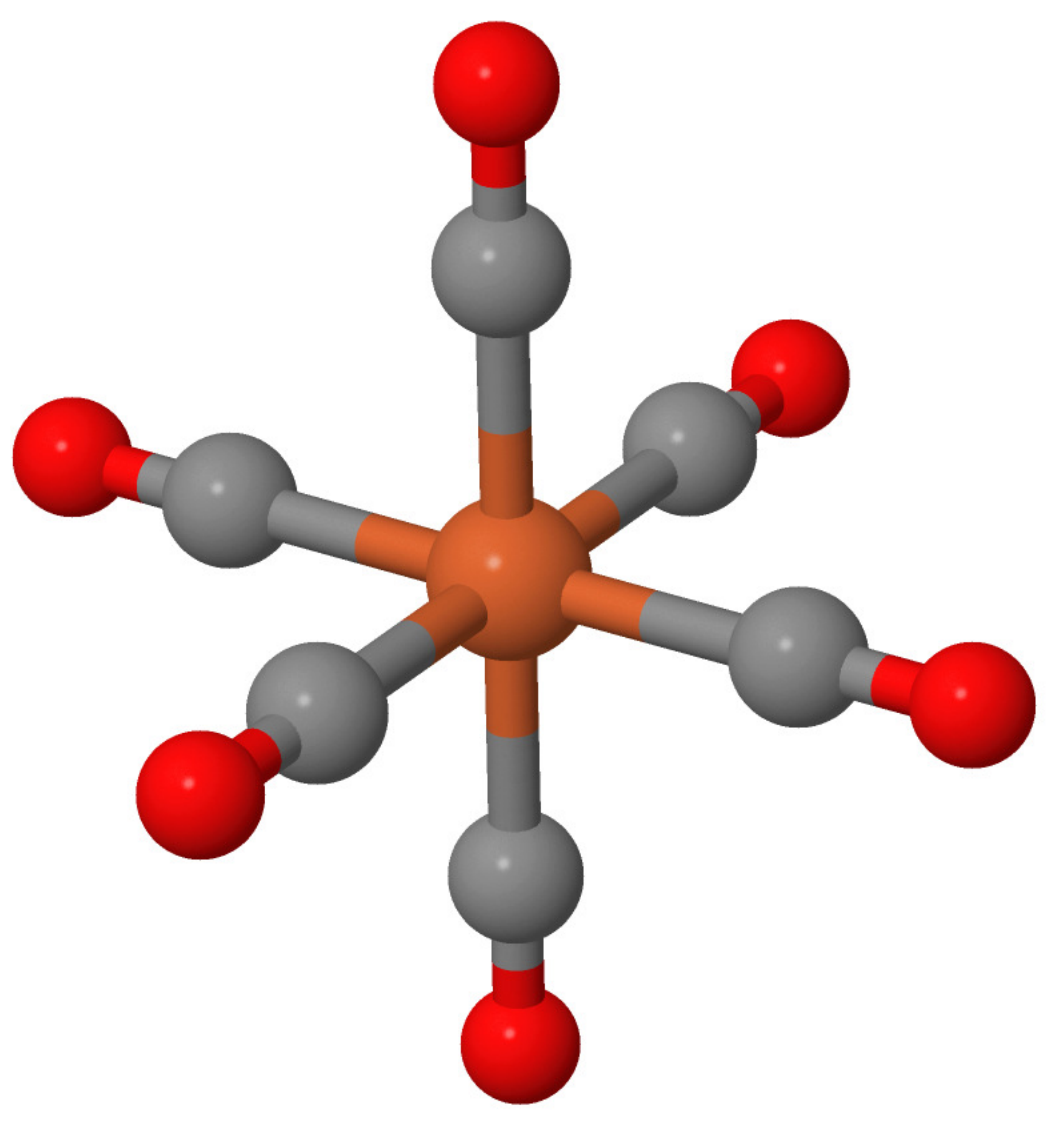}
       \caption{[Fe(CO)$_6$]$^{2+}$}
       \label{fig:FeCO6}
    \end{subfigure}
    \begin{subfigure}[b]{0.2\textwidth}
       \centering
       \includegraphics[width=0.7\textwidth]{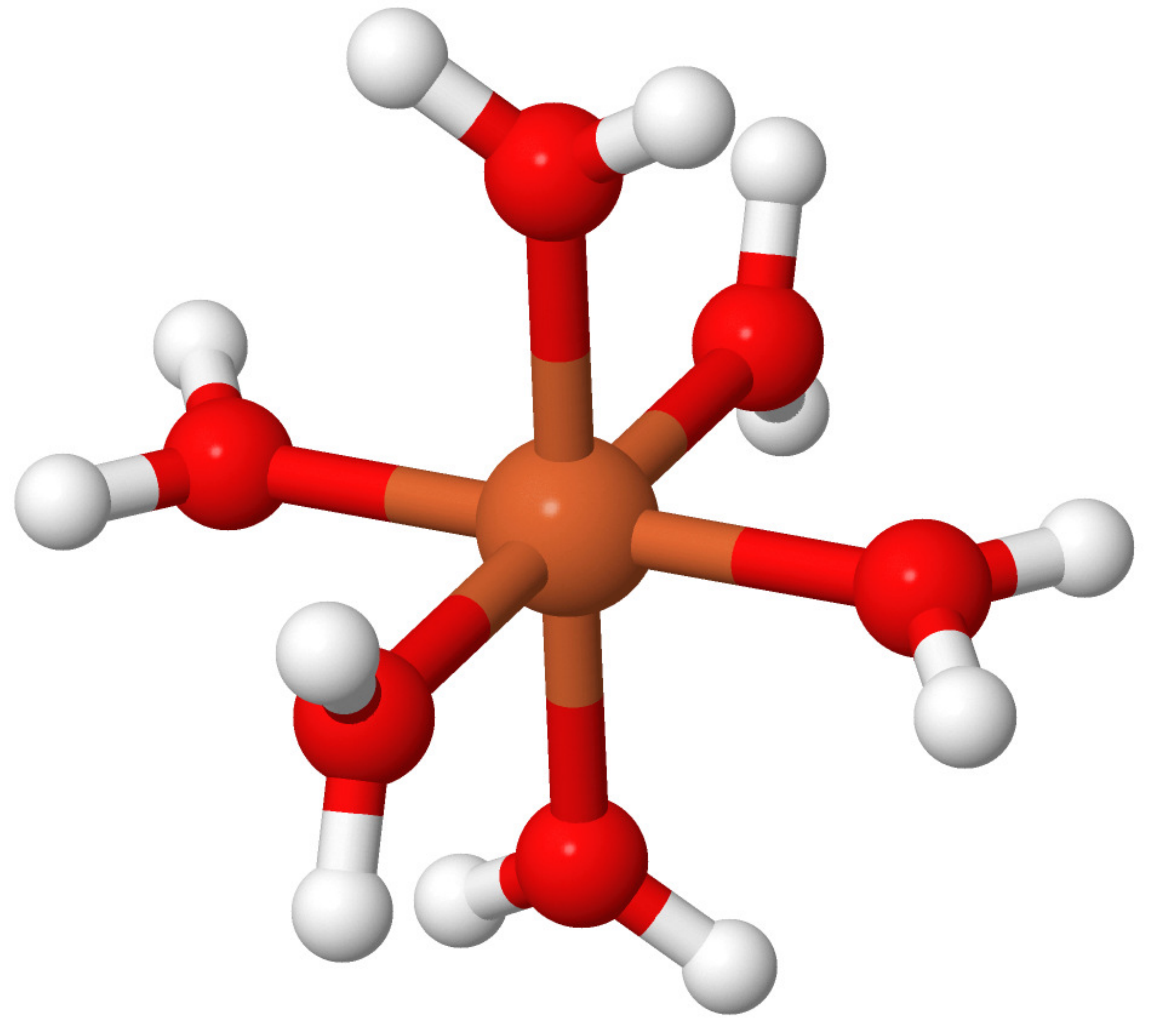}
       \caption{[Fe(H$_2$O)$_6$]$^{2+}$}
       \label{fig:FeH2O}
    \end{subfigure}
    \begin{subfigure}[b]{0.2\textwidth}
       \centering
       \includegraphics[width=0.7\textwidth]{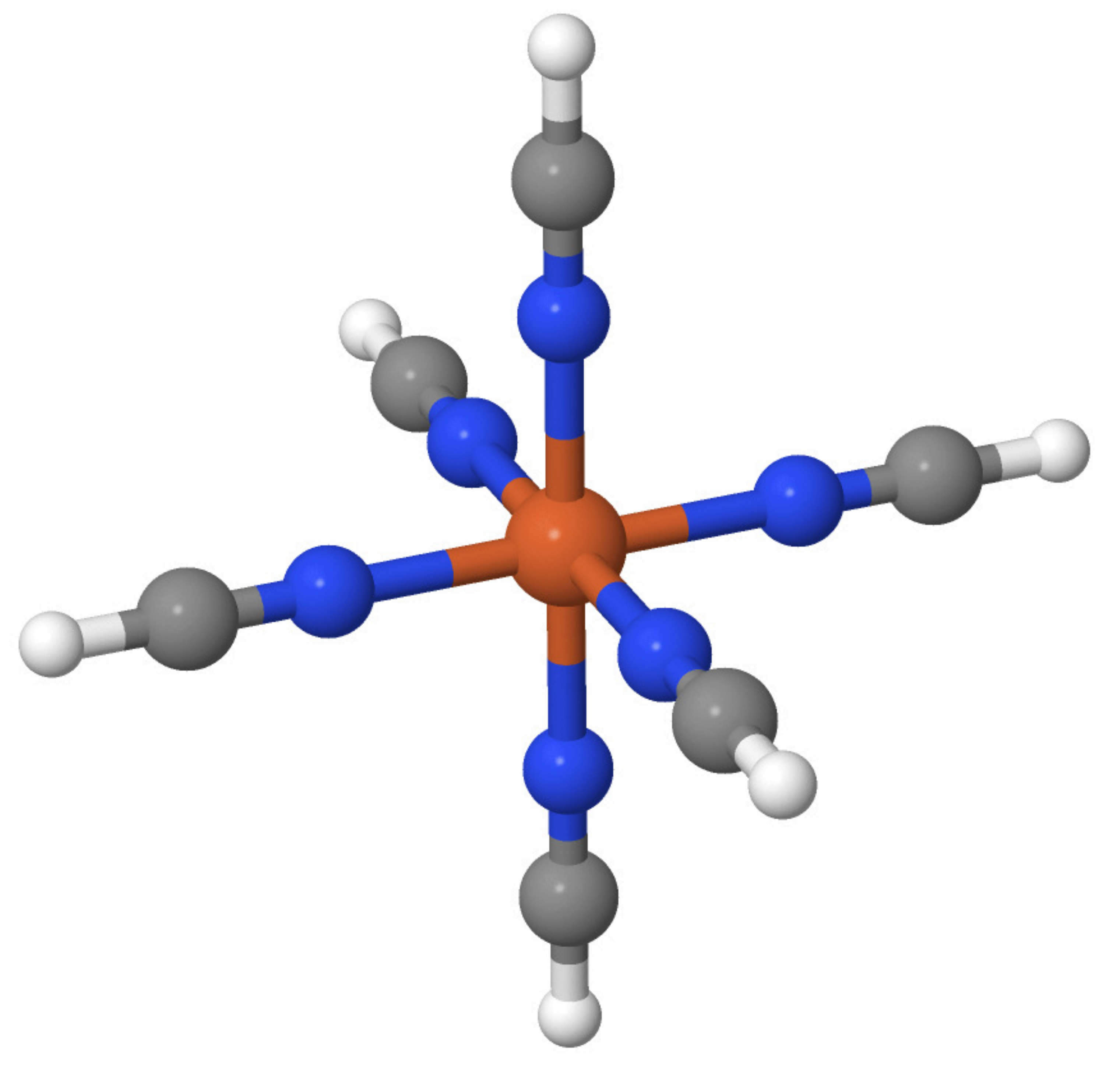}
       \caption{[Fe(NCH)$_6$]$^{2+}$}
       \label{fig:FeNCH}
    \end{subfigure}
        \begin{subfigure}[b]{0.2\textwidth}
       \centering
       \includegraphics[width=0.7\textwidth]{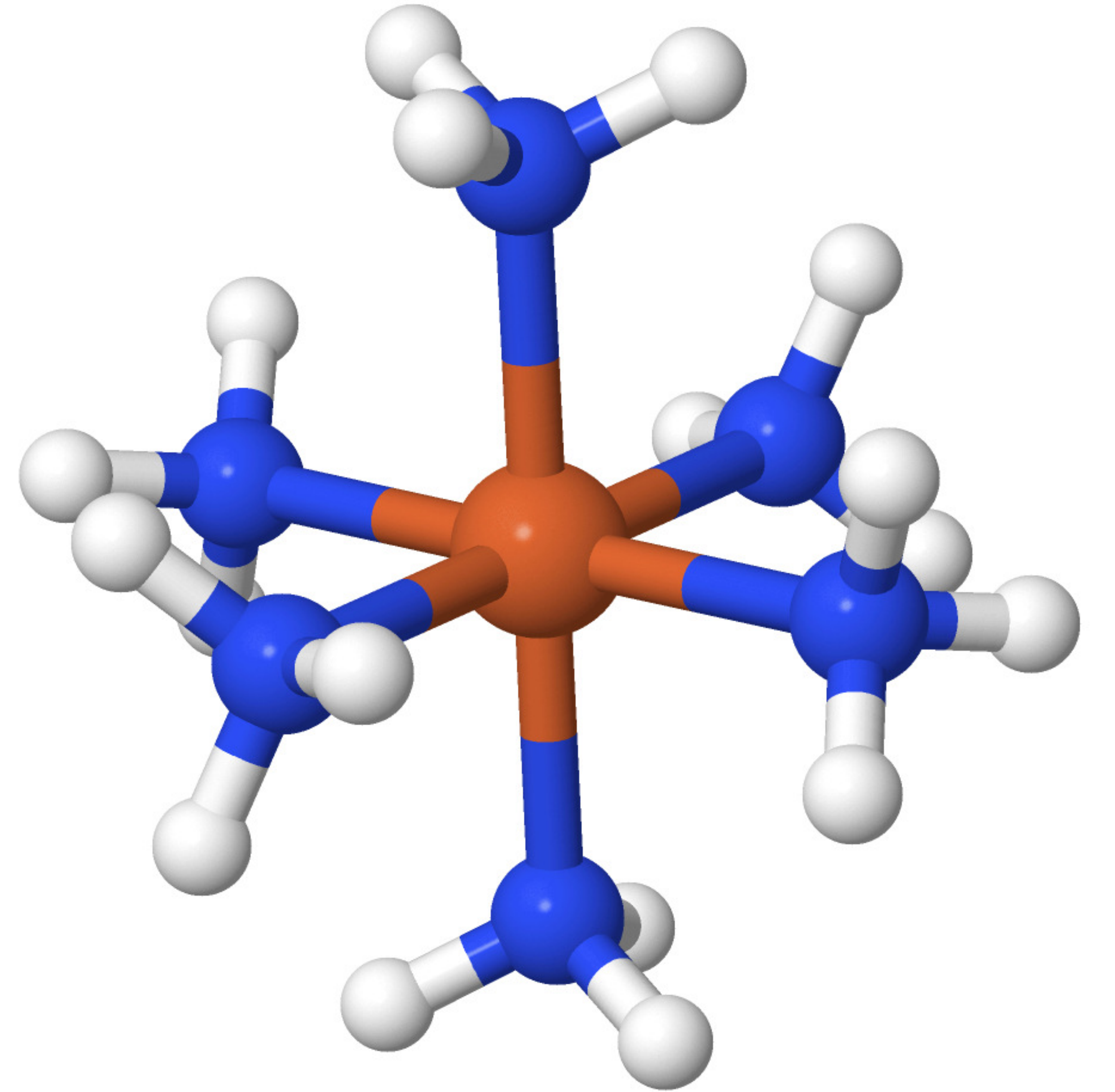}
       \caption{[Fe(NH$_3$)$_6$]$^{2+}$}
       \label{fig:FeNH3}
    \end{subfigure}
    \caption{ The molecular structures for the systems [FeL$_6$]$^{2+}$  for L = CO, H$_2$O, NCH, and NH$_3$.}
    \label{fig:fe_complexes_structure}
\end{figure}

 \section{\label{sec:compdetails} 
 Computational Details}
All of our calculations were performed using the 
FLOSIC code\cite{FLOSICcode,FLOSICcodep} 
at the all-electron level and 
use an extensive Gaussian basis set optimized for the PBE functional.\cite{PhysRevA.60.2840} 
We note that our results for the PBE functional reproduce the literature 
results\cite{wilbraham2018communication} validating 
the choice of present of basis sets.
The LSDA functional as parameterized in the PW92 functional \cite{PhysRevB.45.13244} is used in the SIC-LSDA calculations.
 LSIC is well-defined for the LSDA functional with no gauge dependency \cite{bhattarai2020sdSIC}. 
 Our earlier experience however shows that LSIC often
 performs well with PBE functional \cite{bhattarai2020sdSIC,waterpolarizability} 
 even though the formal gauge dependency occurs. 
 We applied LSIC with both LSDA and PBE functionals in this study.
 The self-consistent FLOSIC calculations with $z_{\sigma}=1$, i.e. PZSIC,  can be 
performed either using optimized effective potential within the 
Krieger-Li-Iafrate approximation \cite{diaz2021kli}
or  using the Jacobi update approach \cite{yang2017flojac}.
Here we used the latter approach.
The quasi self-consistent LSIC (qLSIC) calculations are performed 
by ignoring the variation of scaling factor as explained in Ref. \onlinecite{waterpolarizability}.
In this approach local scaling is applied 
to the SIC potential and to the SIC energy density as shown in  
[Eq. (\ref{eq:lsic})].
We also computed LSIC-DFA total energies using the self-consistent PZSIC densities.
A self-consistency tolerance for the total energy of $10^{-6}$ $E_h$ was used in all calculations. 
The FOD positions were optimized within PZSIC
until the forces\cite{doi:10.1063/1.4907592,PEDERSON2015153}
on the FODs dropped below $10^{-3}$ $E_h$/$a_0$. 
All calculations presented herein employed these FOD positions.
 
 Since the self-consistent FLOSIC calculations on the transition metal 
 complexes are not yet very common and are usually challenging, we briefly outline  our procedure  for reproducibility of our 
 results.
 To initiate self-consistent FLOSIC calculation, one needs not only 
 an initial guess of density but also an initial set of FODs that are used to construct initial FLOs.
 In this work we use various initial guesses for the density
 and FOD sets that are compatible to the initial densities. 
 These initial FOD configurations are generated using our recently developed scheme that 
 generates an FOD structure from any single determinantal wave function.
 We have generated multiple sets of initial FOD configurations using self-consistent 
 densities from various DFAs for each complex.  Additionally, we also begin self-consistent cycle using 
 the superposition of self-interaction-corrected (with LSDA functional) 
 atomic potentials. This procedure usually works.
 In some cases the self-interactions-correction can rapidly change the electron 
 density in the self-consistent cycle resulting in non-compatibility between 
 an initial FOD structure (used at the start of self-consistency)
 and the electron density at a given self-consistent cycle. 
 This breaks the self-consistent FLOSIC cycle due to small eigenvalues
 of overlap matrix during  L\"owdin  orthognalization indicating 
 linear dependence of FLOs. This is one of the major difficulties
 in performing self-consistent FLOSIC calculations.
 In such cases  we use partially self-interaction-corrected densities
 from previous iteration to generate a new set of FODs and restart 
 the calculations. This procedure can be iterated as needed. 
 The details of this method  will be  published elsewhere.
 Starting calculation using
 multiple initial guesses and FOD configurations resulted 
 in the same final energies for all complexes 
 within the tolerances mentioned in the 
 computational details section.  We have included the atomic 
 positions used in this work along with the corresponding final (optimized) 
 set of FOD positions in the supplementary information.

\begin{table*}
    \caption{\label{tab:MAE_vs_DMC_PZSIC} The spin-state gaps (in eV) calculated with LSDA and PBE functional calculated at the DFA, PZSIC, LSIC, and qLSIC levels. MAEs are with respect to diffusion Monte Carlo (DMC). DMC and CCSD(T) values are from Ref. \onlinecite{wilbraham2018communication}. } 

    \begin{tabular*}{0.8\textwidth}{@{\extracolsep{\fill}}lcccccc}
    \toprule
	System	&	{[Fe(H$_2$O)$_6$]}$^{2+}$	&	{[Fe(CO)$_6$]}$^{2+}$	&	{[Fe(NCH)$_6$]}$^{2+}$	&	{[Fe(NH$_3$)$_6$]}$^{2+}$	&	MAE(eV)	\\
	\hline											
	LSDA	&	-0.49	&	5.24	&	2.29	&	0.98	&	2.90	\\
	PBE	&	-1.17	&	3.40	&	0.96	&	0.08	&	1.72	\\
	\hline											
	PZSIC-LSDA	&	1.50	&	2.11	&	1.81	&	2.00	&	2.75	\\
	PZSIC-PBE	&	1.16	&	0.93	&	1.27	&	1.35	&	2.08	\\
	\hline											
	LSIC-LSDA	&	-1.26	&	-0.49	&	-1.15	&	-0.82	&	0.51	\\
	LSIC-PBE	&	-1.06	&	3.40	&	-1.02	&	-0.60	&	1.08	\\
	\hline											
	qLSIC-LSDA	&	-1.08	&	0.40	&	-0.54	&	-0.50	&	0.56	\\
	qLSIC-PBE	&	-0.94	&	4.45	&	-0.44	&	-0.34	&	1.58	\\
	\hline											
	CCSD(T)	&	-1.75	&	1.33	&	-0.43	&	-0.78	&	0.49	\\
	DMC	&	-1.78	&	0.59	&	-1.17	&	-1.23	&		\\
    \bottomrule
    \end{tabular*}
\end{table*}

\section{Results and Discussion}
\label{sec:resultsanddiscussion}
 
 The octahedral Fe(II) complexes studied in this work are
 shown in Fig. \ref{fig:fe_complexes_structure}. 
 The spin state ordering of such complexes has been studied earlier by a number of groups
 to examine the performances of various methods in predicting 
the spin-state gaps.\cite{WOS:000806610000021, wilbraham2018communication, radon2019benchmarking, kulik2020making, nandy2020large, alipour2020appraising, vela2020thermal, mariano2020biased, mariano2021improved, floser2020detailed, song2018benchmarks, patra2019rethinking, droghetti2012assessment, ganzenmuller2005comparison, bowman2012low, fouqueau2004comparison, fouqueau2005comparison, pierloot2006relative, pierloot2008relative, wilbraham2017multiconfiguration, wilbraham2018communication,  verma2017assessment, ye2010accurate, ioannidis2015towards, rudavskyi2014computational, radon2019benchmarking, fumanal2016DMC, pinter2018conceptual, mortensen2015spin, lawson2012accurate, kepenekian2009energetics, domingo2010spin, janesko2017reducing, swart2008accurate}
We use the complexes studied earlier by 
Wilbraham and coworkers\cite{wilbraham2018communication} at the double-hybrid levels. 
These complexes are in octahedral conformation with four different ligands (L) where L=H$_2$O, NH$_3$, NCH, and CO. 
The ligand field strengths of these four ligands are different.
 The  hexa-aqua and hexa-amine chelated complexes  
 are examples of weak-field limits while the   hexa-carbonyl represents 
 a strong field limit.
We used the geometries of Wilbraham and coworkers\cite{wilbraham2018communication} which are optimized 
 at the PBE0\cite{adamo1999toward}  level of theory, using the modified def2-TZVPP\cite{weigend2006accurate} large basis set.
 The valence electronic configuration of Fe is $3d^6 ~4s^2$. 
 The spin states considered for the Fe(II) complexes  are singlet (low spin 
 state) 
 and the quintet (high spin state).
 In the present calculations the spin-gap
 between the quintet (HS) and singlet (LS)
 is defined as
 \begin{equation}\label{eq:spingap}
  \Delta E^{HS-LS} = E(HS) - E(LS).
 \end{equation}

\begin{table*}
    \caption{\label{tab:MAE_vs_DMC_DFA} The spin-state gap values (in eV)
    for [FeL$_6$]$^{2+}$ calculated with LSDA and PBE 
    from the DFA part in PZSIC and qLSIC calculations (that is, the first term on right hand side of Eq. (\ref{eq:spingap_dfa}).
    MAEs are computed against DMC. DMC and CCSD(T) reference values are from Ref. 
    \onlinecite{wilbraham2018communication}. 
    }
    \begin{tabular*}{1.0\textwidth}{@{\extracolsep{\fill}}ccccccc}
    \toprule
	System	&	{[Fe(H$_2$O)$_6$]}$^{2+}$	&	{[Fe(CO)$_6$]}$^{2+}$	&	{[Fe(NCH)$_6$]}$^{2+}$	&	{[Fe(NH$_3$)$_6$]}$^{2+}$	&	MAE(eV)	\\
	\midrule											
	$\Delta E^{PZSIC-LSDA}_{LSDA}$	&	-1.06	&	2.55	&	0.28	&	-0.17	&	1.30	\\
	$\Delta E^{PZSIC-LSDA}_{PBE}$	&	-1.55	&	0.93	&	-0.76	&	-0.86	&	0.34	\\

	\midrule											

	$\Delta E^{PZSIC-PBE}_{LSDA}$ 	&	-1.08	&	2.31	&	0.11	&	-0.20	&	1.18	\\

	$\Delta E^{PZSIC-PBE}_{PBE}$ 	&	-1.56	&	0.69	&	-0.92	&	-0.88	&	0.23	\\
	\midrule									
	 $\Delta E^{qLSIC-LSDA}_{LSDA}$&	-0.67	&	4.71	&	1.69	&	0.59	&	2.48	\\
    $\Delta E^{qLSIC-PBE}_{PBE}$ &	-1.29	&	2.99	&	0.42	&	-0.24	&	1.37	\\	
	\midrule
	CCSD(T)	&	-1.75	&	1.33	&	-0.43	&	-0.78	&	0.49	\\
	DMC	&	-1.78	&	0.59	&	-1.17	&	-1.23	&		\\
    \bottomrule
    \end{tabular*}
\end{table*}

The spin-state gaps defined as $\Delta E^{HS-LS}$ are presented for the LSDA, PBE, 
PZSIC-LSDA, and PZSIC-PBE in Table \ref{tab:MAE_vs_DMC_PZSIC}.
The diffusion Monte-Carlo (DMC) 
spin-gap energy values from Ref. \onlinecite{song2018benchmarks} are taken as reference. 
For the 
sake of comparison, coupled-cluster single double and perturbative triple [CCSD(T)] 
values are also  presented.
The positive (negative) values in the Table \Ref{tab:MAE_vs_DMC_PZSIC} indicate 
that the LS (HS) state is more stabilized than HS (LS) state. 
 First, we would like to note that our PBE results are essentially identical to those of 
 Wilbraham and coworkers\cite{wilbraham2018communication} validating our choices
 of computational parameters and especially of the basis sets.
 The DFAs, both LSDA and PBE,  predict the sign of the 
spin-state gaps correctly for a weak field ligand H$_2$O and for a 
strong field ligand $\mathrm{CO}$.  
 However, the errors for the spin-gaps are rather large, especially for $\mathrm{CO}$.
 For the other two ligands, 
 both LSDA and PBE favor low spin states. 
  This is generally 
 known\cite{wilbraham2018communication, mariano2020biased,mariano2021improved}
 and has been attributed due to delocalization errors 
 (or self-interaction errors) of these functionals.
 On the other hand, the  Hartree-Fock method which is 
 one electron self-interaction-free and lacks dynamical
 correlation tend to stabilize high spin states over low spin 
 states. This observation has led to a few studies that 
 explored the reduction of over stabilization by DFAs 
 by mixing various percent of HF exchange in the global
 hybrid functionals\cite{ioannidis2015towards,pinter2018conceptual,wilbraham2018communication}. In general, global hybrids
 functional perform better than the local and semi-local 
 functionals.
 It has also been noted by
 Song \textit{et al.} that  using  HF density in the semi-local 
 functional results in improved prediction of the 
 spin-state gaps by the semi-local functionals\cite{song2018benchmarks}.
 This observation suggests that one-electron self-interaction correction 
 methods such as PZSIC or LSIC might improve spin-gaps. 
 
    The spin-state gaps predicted by the PZSIC method are included in Table  \ref{tab:MAE_vs_DMC_PZSIC}. 
   It is evident that PZSIC method does not improve upon the bare DFA functionals except for the CO ligand.
    The mean absolute errors (MAEs) with LSDA, PBE, PZSIC-LSDA, 
   PZSIC-PBE are 2.90, 1.72, 2.75, and 2.08 eV, respectively. The MAEs
   show that for the PZSIC-LSDA barely improves LSDA results while for 
   the PBE functional PZSIC performs worse than uncorrected PBE functional.
   A closer look at the results indicates that PZSIC has a tendency to stabilize the low 
   spin states.
   Even in the case of  a weak field ligand H$_2$O,  where uncorrected
   LSDA and PBE functionals give qualitatively correct results,
   the PZSIC favors low spin states 
   thus worsening the 
   DFA spin gap results qualitatively and quantitatively.
   It has been known that the PZSIC does not improve upon DFAs (especially  semi-local DFAs)
   for thermochemical properties but its performance for spin-state gaps has not been known.
   This failure of the PZSIC method to predict correct spin states is
   rather surprising and is the most unexpected result of this work.
   We will return to  the discussion of failure of PZSIC method 
   towards the end of this section.

   The scaling down of the SIC in the many-electron regions using the LSIC method brings the
   results closer to the DMC results (Cf.  Table  \ref{tab:MAE_vs_DMC_PZSIC}).
   This is a perturbative approach that does not change the PZSIC density.
   The one-shot LSIC method gives results that are in more 
   quantitative  agreement with the reference values. 
   The MAE for LSIC-LSDA and LSIC-PBE drops to 0.51 and 
   1.08 eV from 2.75 and 2.08 eV for PZSIC-LSDA and PZSIC-PBE functionals, 
   respectively.
   For comparison, the CCSD(T) MAE (0.49 eV) is comparable to LSIC-LSDA MAE.
   Since one-shot LSIC is a perturbative approach that does not change the PZSIC density,
   the improvement comes  directly from the LSIC energy functional.
   The perturbative LSIC significantly reduces the excessive SIC correction of the PZSIC 
   thereby improving the results but it slightly overcorrects in case of the strong ligand
   $\mathrm{CO}$.
   On the other hand, 
   the quasi-LSIC (qLSIC), where in the orbital dependent SIC potentials are scaled, correctly predicts 
   the sign of spin-state gaps for all the complexes studied here. The MAE of the spin-gaps in the case of qLSIC-LSDA is 0.56 eV which is comparable to that of CCSD(T) (0.49 eV).
   The results also show that both  qLSIC and one shot LSIC perform better with the simpler LSDA than with the PBE functional.

 To understand the  improved performance of the LSIC method and 
 the failure of PZSIC method,  we analyze the contributions  
 to the spin-state gap from the DFA energy terms, i.e., the first term on the right hand side (RHS) of Eq. (\ref{eq:lsic}), and 
 the sum of orbitalwise SIC corrections (i.e., Eq. (\ref{eq:lsic_esic})).
 Thus we substitute Eq.(\ref{eq:lsic}) into the Eq. (\ref{eq:spingap}) resulting in the following expression.
 \begin{equation}
     \Delta E^{HS-LS}=\Delta E^{HS-LS}_{DFA} + \Delta E^{HS-LS}_{SIC}.
     \label{eq:spingap_dfa}
 \end{equation}
 The first term on the RHS is therefore  the DFA spin-state gap evaluated using the 
 SIC density and the second term is the  correction due to the energy.  
 This analysis, in some sense, is similar to the 
 HF-DFT approach\cite{VERMA201210,PhysRevLett.111.073003,vuckovic2019density,janesko2021replacing}. 
 Unlike the HF-DFT however, the advantage here is  that the $E^{DFA}$ contributions are part 
 of the PZSIC and qLSIC energies and no additional calculations are required.

\begin{figure}
    \centering
    \begin{subfigure}[b]{0.22\textwidth}
       \centering
       \includegraphics[width=0.9\textwidth]{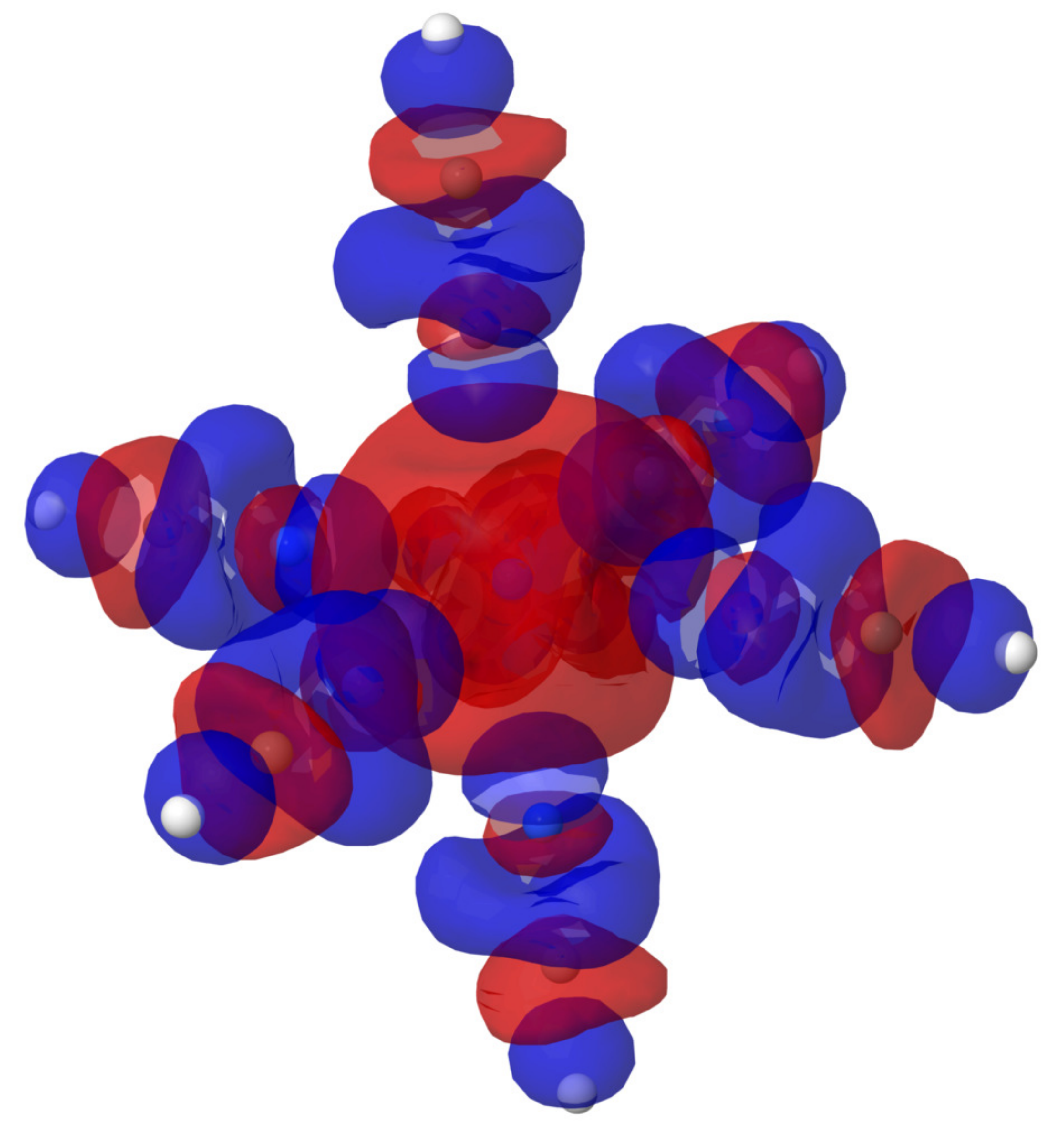}       \caption{PZSIC-LSDA - LSDA density difference.}
       \label{fig:nch_ls_pzsic_dfa}
    \end{subfigure}
    \begin{subfigure}[b]{0.22\textwidth}
       \centering
       \includegraphics[width=0.9\textwidth]{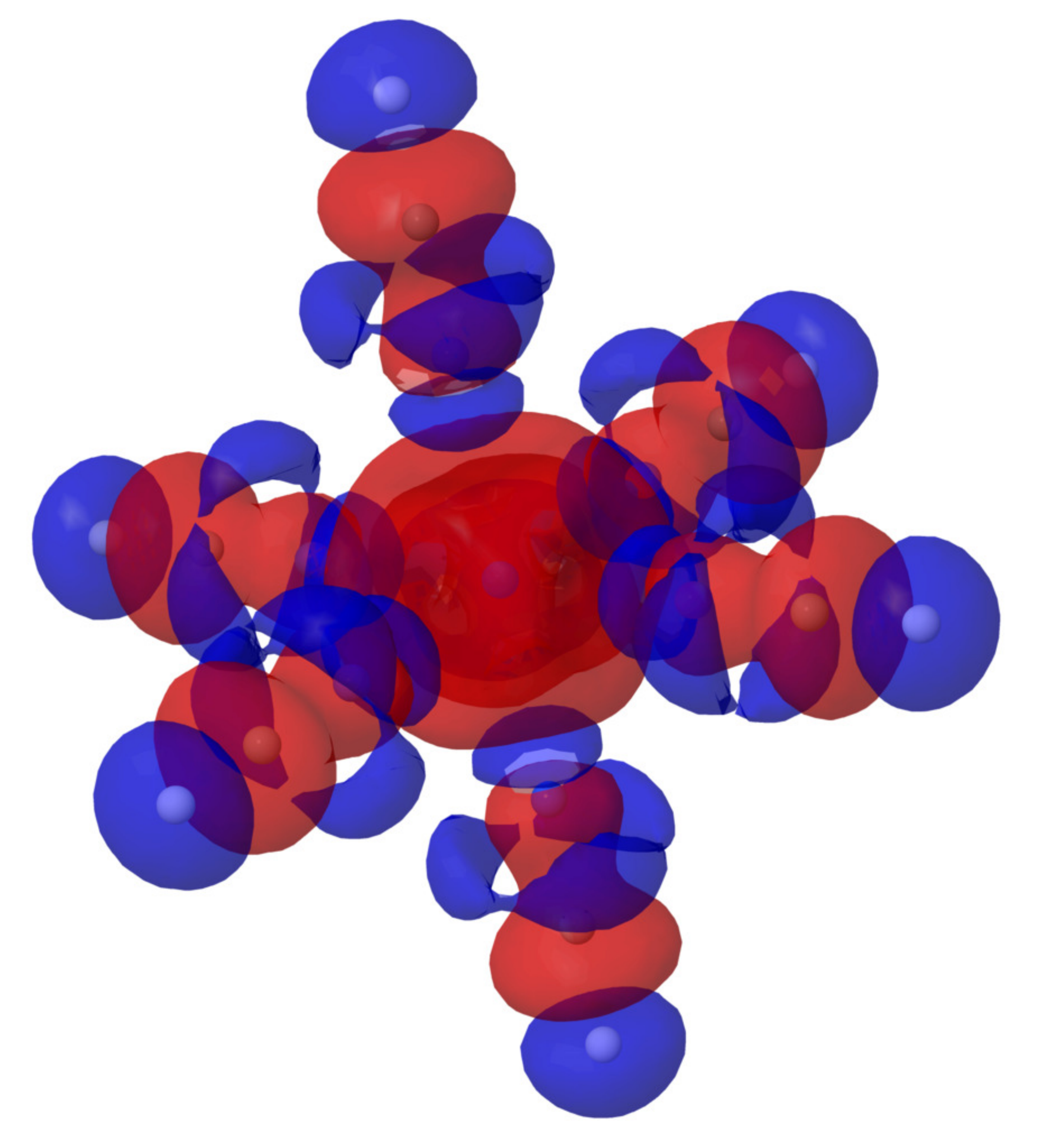}
       \caption{qLSIC-LSDA - LSDA density difference. }
       \label{fig:nch_ls_qlsic_dfa}
    \end{subfigure}
    \caption{Total density difference of (a) PZSIC-LSDA and LSDA and (b) qLSIC-LSDA and LSDA for the [Fe(NCH)$_6$)]$^{2+}$ complex in the LS state. 
    The isosurface value used for both images is 0.0005e and the red (blue) surface shows regions where LSDA density is larger (smaller). }
    \label{fig:nch_ls_densities_plot}
\end{figure}

 The DFA contribution to the spin-state gaps calculated on self-consistent PZSIC-DFA and qLSIC-DFA densities 
   are shown in Table~\ref{tab:MAE_vs_DMC_DFA}. To simplify the discussion below 
   we use $\Delta E^{Den}_{En}$ notation to convey the calculation scheme where the subscript refers to the 
   energy functional and the superscript refers to the underlying functional for the SIC 
   density.
   The MAE of  $\Delta E^{PZSIC-LSDA}_{LSDA}$ is 1.30 eV. This is significantly better than that 
   of LSDA (MAE = 2.90 eV). Thus, the density changes due to SIC (to the potential)
   within the PZSIC method result in reduction of MAE by about 1.6 eV. 
   Even more spectacular  reduction in error is  obtained in case of PBE functional.
    The $\Delta E^{PZSIC-PBE}_{PBE}$ has MAE of only 0.23 eV in contrast to 2.08 eV of
   of PZSIC-PBE. This is half the MAE of the CCSD(T) method (0.49 eV).
   As all self-consistent calculations performed with SIC methods herein generate 
   {\it self-interaction free} densities, we can use these SIC densities in uncorrected 
   DFAs used in this work to get some idea about error cancellations that play role in predicting
   spin-state gaps. It turns out that 
   the PBE functional performs excellently not only with the PZSIC-PBE density 
   but also with PZSIC-LSDA densities (with MAE = 0.34 eV).
   This suggests that  more sophisticated meta-GGA functionals may also perform well for spin-state
   gaps but that discussion is beyond the scope of present work.
   The use of qLSIC densities in the uncorrected DFAs however do not show similar level of 
   improvement in the spin state gaps.
   For example, the $\Delta E^{qLSIC-LSDA}_{LSDA}$ has an MAE of 2.48 eV which is comparable
   to that of LSDA. Similarly, the $\Delta E^{qLSIC-PBE}_{PBE}$ has an MAE of 1.37 eV. 
   These results are not very surprising. It is apparent from Eq. (\ref{eq:spingap_dfa}) that 
   any method that performs well with $\Delta E^{HS-LS}_{DFA}$ 
   for spin-state gaps must yield  a small contribution from the SIC term  ($ \Delta E^{HS-LS}_{SIC}$).
   This gives a hint that a poor performance of PZSIC is therefore due to its excessive SIC to the energy. 
   Indeed such a tendency of PZSIC to excessively correct has also been seen in earlier studies.\cite{doi:10.1063/1.1794633,waterpolarizability,zope2019LSIC,doi:10.1063/1.5120532}
   This excessive correcting tendency of PZSIC results in  somewhat greater localization of density which
   might be  beneficial for use in uncorrected DFAs for removing delocalization errors as done 
   in the HF-DFT or density corrected DFT and also seen in the present results.  As discussed below, among
   all the self-consistent SIC methods 
   used here the PZSIC density is the most localized.
   The density differences between the PZSIC and LSDA, 
   and between qLSIC and LSDA are plotted in Fig. \ref{fig:nch_ls_densities_plot} for Fe(NCH)$_6^{2+}$. The red isosurface
   shows the regions where DFA density is larger than the qLSIC/PZSIC density. The LSDA density 
   is higher in the interatomic region around the Fe center compared to the SIC density. This 
   plot also shows that although the density differences with PZSIC and qLSIC are similar,
   with qLSIC the differences are smaller particularly near the Fe center. 
   To obtain additional evidence about the (greater) localizing tendency of PZSIC, we compute 
   the spin charges at Fe(II) in the complex by integrating the spin densities within 
   atomic spheres centered at Fe and with radius equal to van der Wall radius of Fe. 
   Likewise, we also compute spin charges
   using L\"owdin's population analysis. These results are presented in  Table \ref{tab:population}.
   It is evident from the the table that the uncorrected LSDA and PBE  functionals
   have the smallest values of spin charges consistent with a known 
   delocalization  of density within these methods. On the 
   other hand, PZSIC has the largest values of spin charges among all methods suggesting
   its spin density to be most localized amongst all methods. This is consistent with 
   above discussion. On the other hand, the qLSIC  spin charges are intermediate 
   between the PZSIC and uncorrected DFA spin charges. The less localized qLSIC density
   is not favorable for use in uncorrected DFAs for the density corrected DFT calculations.
    The qLSIC-LSDA provides most balanced
   description of the spin-state gaps [comparable to CCSD(T)] where in both the terms in Eq. 
   (\ref{eq:spingap}) contribute to the gap. The qLSIC-LSDA also predicts the correct sign for spin-state
   gaps for all ligands. The qLSIC-PBE doens't perform as well as qLSIC-LSDA possibly due 
   to gauge-dependence\cite{bhattarai2020sdSIC}.
   
   \begin{table*}
    \caption{\label{tab:population}
    Spin charges (in e) at the Fe site for the high spin state in various approximations.   Spin charges are obtained from L\"owdin population 
    analysis and by integrating spin density using atomic sphere (see text for details).}
    \begin{tabular*}{1.0\textwidth}{@{\extracolsep{\fill}}lcccccccc}
    \toprule
		&		&	\multicolumn{2}{c}{H$_2$O}	& \multicolumn{2}{c}{}	& \multicolumn{2}{c}{CO}		&		\\
	   \cmidrule(lr){2-5} \cmidrule(lr){6-9}
		&	\multicolumn{2}{c}{LSDA}		&	\multicolumn{2}{c}{PBE}		&	\multicolumn{2}{c}{LSDA}		&	\multicolumn{2}{c}{PBE}		\\
	\cmidrule(lr){2-3} \cmidrule(lr){4-5} \cmidrule(lr){6-7} \cmidrule(lr){8-9}
	Population analysis	&	Atomic sphere	&	L\"owdin	&	Atomic sphere	&	L\"owdin	&	Atomic sphere	&	L\"owdin	&	Atomic sphere	&	L\"owdin	\\  \hline
	DFA	&	3.607	&	3.648	&	3.637	&	3.674	&	3.509	&	3.665	&	3.554	&	3.695	\\ 
	PZSIC 	&	3.768	&	3.799	&	3.764	&	3.793	&	3.734	&	3.792	&	3.708	&	3.797	\\ 
	qLSIC 	&	3.686	&	3.724	&	3.703	&	3.737	&	3.587	&	3.715	&	3.608	&	3.727	\\
    \bottomrule
    \end{tabular*}
\end{table*}

\section{Conclusions}
In summary,  we  have examined the effect of 
SIC on the spin-state gaps of octahedral [Fe(II)L$_6$]$^{2+}$ complexes  with L=CO, NCH, NH$_3$, and H$_2$O
using the LSIC and PZSIC methods.
One surprising result is that the PZSIC 
fails to improve upon  
the DFA results for weak field ligands and has a propensity to stabilize low spin states.
Analysis of results show that the  PZSIC energy functional provides too 
much SI correction to the energy. In fact,
no PZSIC energy correction is needed as DFA part of PZSIC functional alone is sufficient to provide 
good spin-gaps. The analysis of results show that the self- PZSIC produces
more localized density than the quasi-self-consistent LSIC method. The more 
localized density is favorable for use in uncorrected functional 
as in the HF-DFT or density corrected DFT. Indeed the spin-state gaps 
obtained using DFA functional on PZSIC  densities are significantly improved compared 
to those of self-consistent PZSIC  or DFA results.
This improvement is spectacular for the PBE functional.
The perturbative LSIC  results evaluated on the PZSIC 
densities predicts accurate spin-state gaps and its performance is comparable to that 
of CCSD(T) method even with the simplest DFA.  Likewise, qLSIC-LSDA predicts correct sign
for spin-gaps for all ligands with mean absolute error comparable to that of CCSD(T).
LSIC does not work that well with PBE functional especially for the CO ligand possibly 
due to gauge  dependency.
Still, it is very gratifying that the simplest (LSDA) density functional when corrected for 
SIE using the LSIC method, accurately predicts
spin-state gaps of Fe(II) complexes and predicts correct spin states as most 
stable states.
It remains unclear if  the LSIC method 
will perform to the same degree of  excellence for other 
SCO complexes. We will be undertaking a large scale study to address this question.

\section*{Data Availability Statement}
The data that supports the findings of this study are available within the article.     

\begin{acknowledgments}
Authors acknowledge  Dr. Yoh Yamamoto for discussions.
This work was supported by the US Department of Energy, Office of
Science, Office of Basic Energy Sciences, as part of the 
Computational Chemical Sciences Program under Award No. 
DE-SC0018331. Support for computational time at the Texas Advanced 
Computing Center 
and at NERSC is gratefully acknowledged.
SR expresses his gratitude to the Mexican National Council for Science and Technology (CONACYT)
for financial support.
\end{acknowledgments}

\bibliography{bibtex2}
\end{document}